\documentclass[useAMS, onecolumn]{mn2e}
\usepackage{epsfig}
\usepackage{amsmath}
\usepackage{multirow}
\def\be{\begin{equation}}
\def\ee{\end{equation}}
\def\ba{\begin{eqnarray}}
\def\ea{\end{eqnarray}}
\def\go{\mathrel{\raise.3ex\hbox{$>$}\mkern-14mu
             \lower0.6ex\hbox{$\sim$}}}
\def\lo{\mathrel{\raise.3ex\hbox{$<$}\mkern-14mu
             \lower0.6ex\hbox{$\sim$}}}

\def\tomega{{\tilde{\omega}}}

\begin{document}
\title[Inertial-Acoustic Modes in Black-Hole Accretion Discs]
{Corotational Instability of Inertial-Acoustic Modes in Black-Hole Accretion 
Discs: Non-Barotropic Flows}
\author[D. Tsang and D. Lai]
{David Tsang$^{1,2}$\footnotemark[1] and Dong Lai$^{1}$\thanks{Email:
dtsang@astro.cornell.edu; dong@astro.cornell.edu} \\ 
$^1$Department of Astronomy, Cornell University, Ithaca, NY 14853, USA \\
$^2$Department of Physics,
Cornell University, Ithaca, NY 14853, USA \\}

\label{firstpage}
\maketitle

\begin{abstract}
We study the effect of corotation resonance on the inertial-acoustic
oscillations (p-modes) of black-hole accretion discs. Previous works
have shown that for barotropic flows (where the pressure depends only
on the density), wave absorption at the corotation resonance can lead to
mode growth when the disc vortensity, $\zeta=\kappa^2/(2\Omega\Sigma)$
(where $\Omega,~\kappa,~\Sigma$ are the rotation rate, radial
epicyclic frequency and surface density of the disc, respectively),
has a positive gradient at the corotation radius. Here we generalize
the analysis of the corotation resonance effect to non-barotropic fluids. 
We show that the mode instability criterion is modified by the finite radial
Brunt-V\"as\"al\"a frequency of the disc.  We derive an analytic
expression for the reflectivity when a density wave impinges upon the
corotation barrier, and calculate the frequencies and growth rates of
global p-modes for disc models with various $\alpha$-viscosity
parameterizations.  We find that for disc fluids with constant
adiabatic index $\Gamma$, super-reflection and mode growth depend on
the gradient of the effective vortensity, $\zeta_{\rm eff} =
\zeta/S^{2/\Gamma}$ (where $S \equiv P/\Sigma^{\Gamma}$ measures the
entropy): when $d\zeta_{\rm eff}/dr > 0$ at the corotation radius,
wave absorption leads to amplification of the p-mode.  Our
calculations show that the lowest-order p-modes with azimuthal wave
number $m=2,~3,~4,\ldots$ have the largest growth rates, with the
frequencies approximately in (but distinct from) the $2:3:4\ldots$
commensurate ratios. We discuss the implications of our results for
the high-frequency quasi-periodic oscillations observed in accreting
black-hole systems.
\end{abstract}

\section{Introduction}

High frequency quasi-periodic oscillations (QPOs) in X-ray binary
systems have been observed for a number of years and may provide an
important tool for studying the strong gravitational fields of black
holes (see Remillard \& McClintock 2006). However, the physical
mechanisms that generate such X-ray variability remain unclear. One of
the most appealing models for the source of QPOs is the relativistic
diskoseismic oscillation model, where general relativistic effects
produce trapped oscillation modes at the inner region of an accretion
disc (e.g., Kato \& Fukue 1980; Okazaki et al. 1987; Nowak \& Wagoner
1991; see Wagoner 1999 and Kato 2001 for reviews). 
Other related works on black-hole diskoseismology, such as possible mode
excitation and damping (e.g., Ortega-Rodriguez \& Wagoner 2000; 
Li, Goodman \& Narayan 2003; Kato 2003,2008; 
Tagger \& Varniere 2006; Ferreira \& Ogilvie 2009; Tsang \& Lai 2009a), 
the effects of disc magnetic fields (e.g., Tagger \& Pellat 1999; 
Fu \& Lai 2009) and numerical simulations (e.g., Arras et al.~2006; 
Reynolds \& Miller 2008; O'Neill, Reynolds \& Miller 2009), as well as 
other ideas for high-frequecy QPOs, such as non-linear resonances (e.g.
Abramowicz \& Kluzniak 1999; Horak \& Karas 2006; Rebusco 2008)
and boundary layer oscillations (e.g. Li \& Narayan 2004; Tsang \& Lai 2009b),
are reviewed in section 1 of Lai \& Tsang (2009).

In Lai \& Tsang (2009), we studied the global corotational
instability of non-axisymmetric
p-modes (also called intertial-acoustic modes) trapped in the
inner-most region of the accretion disc around a black hole. These
modes do not have nodes in the vertical direction, and were shown to
be amplified by the effect of wave absorption at corotation resonance. 
Near the black hole the radial epicyclic frequency $\kappa$ reaches a maximum
and goes to zero at the innermost stable circular orbit (ISCO). This
causes a non-monotonic behavior in the fluid vortensity, $\zeta =
\kappa^2/(2\Sigma\Omega)$, such that $d\zeta/dr > 0$ inside the radius
where $\zeta$ peaks. It can be shown that the sign of the corotational
wave absorption depends on the sign of the vortensity gradient $d\zeta/dr$
(Tsang \& Lai 2008; see Goldreich \& Tremaine 1979). Thus 
p-modes with positive vortensity gradient at the corotation radius can be
overstable due to corotational wave absorption. 
Tagger \& Pellat (2002) and Tagger \& Varniere (2006) showed that 
the global p-mode intability can be enhanced when the 
disc is threaded by a strong (of order equipartion), large-scale
poloidal magnetic field.

Our previous study (Lai \& Tsang 2009) and much of the related work
on disc dynamics have assumed barotropic flows for the disc
(i.e. the pressure depends only on density).
This assumption provides convenient simplification, but may miss 
important effects of the disc dynamics. 
For example, Lovelace et al. (1999) and Li et al. (2000) studied the 
adiabatic perturbations for waves trapped by a disk entropy radial profile 
that has a localized maximum, leading to the so-called Rossby-wave
instability. They showed that in such a case the key parameter
determining the effect of the corotation is no longer the gradient of the
vortensity, but rather the slope of a modified effective vortensity, 
$\zeta_{\rm eff} \equiv \zeta/S^{2/\Gamma}$, where 
$S\equiv P/\Sigma^{\Gamma}$ is defined as the entropy and $\Gamma$ 
is the 2-dimensional adiabatic index (assumed to be constant).
As another example, Baruteau \& Masset (2008) showed that the
corotation torque of a protoplanetary disc on a planet can be 
significantly different for barotropic and non-barotropic 
fluids.


In this paper we study the global corotational instability of
p-modes in accretion discs around black holes,
generalizing our pevious works (Tsang \& Lai 2008; Lai \& Tsang 2009)
to include non-barotropic effects.
In section 2 we develop the basic equations of adiabatic perturbations 
for generic accretion discs. In section 3 we analyze the
effect of the corotation resonance, including a careful treatment of
both the first and second-order singularities of the resonance; 
we derive a WKB expression for the reflectivity due to the corotation 
barrier and show that super-reflection can be achieved under
certain conditions.
In section 4 we consider black hole disc models parametrized by 
$\alpha$-viscosity, and calculate the global disc p-mode frequencies
and growth rates. Finally, in section 5 we discuss the implications of this 
work for models of high-frequency QPOs.

\section{Basic Equations}

We begin by considering the basic fluid equations of a 2-dimensional disc. 
The continuity and momentum equations read:
\begin{subequations}
\ba
 \partial_t \Sigma + {\bf u} \cdot {\bf \nabla} \Sigma  + \Sigma {\bf \nabla} \cdot {\bf u} &=& 0~, \\
\partial_t {\bf u} + ({\bf u} \cdot {\bf \nabla}) {\bf u} &=& -\frac{1}{\Sigma} {\bf \nabla} P - {\bf \nabla} \Phi~, 
\ea
\end{subequations}
where $P(r) = \int p\, dz$ is the vertically integrated pressure, and $\Sigma(r) = \int \rho\, dz$ is the surface density and we adopt the Pacyznski-Wiita psuedo-Newtonian potential $\Phi = GM/(r-2r_g)$ with $r_g = GM/c^2$.  
Assuming that the background flow has ${\bf u} = r\Omega \hat{\phi}$, and that the Eulerian perturbations $\delta \Sigma$, $\delta P$, and $\delta {\bf u} = \delta u_r \hat{r} + \delta u_\phi \hat{\phi}$, have the form $\exp(im\phi - i\omega t)$, we find the linear perturbation equations:
\begin{subequations}
\ba
-i \tomega \delta \Sigma + \frac{1}{r}\frac{\partial}{\partial r} ( \Sigma r \delta u_r) + \frac{im}{r}\Sigma \delta u_\phi &=& 0~,\label{perturb1}\\
-i \tomega \delta u_r - 2\Omega \delta u_\phi &=& -\frac{1}{\Sigma} \frac{\partial}{\partial r} \delta P + \frac{\delta \Sigma}{\Sigma^2} \frac{\partial P}{\partial r}~,\\
-i\tomega \delta u_\phi + \frac{\kappa^2}{2\Omega} \delta u_r &=& -\frac{im}{r}\frac{\delta P}{\Sigma}~\label{perturb3},\ea
\end{subequations}
where $c_s^2 = \partial P/\partial \Sigma$ is the adiabatic sound speed, $\tomega = \omega - m\Omega$ and $\kappa$ is the radial epicyclic (angular) frequency.

Tsang \& Lai (2008) and Lai \& Tsang (2009) assumed the disc fluid is barotropic such that $P = P(\Sigma)$. Here we consider adiabatic perturbations of a general non-barotropic disc. The Lagrangian density perturbation $\Delta \Sigma$ and pressure perturbation $\Delta P$ are related by
\be
\Delta \Sigma = \frac{1}{c_s^2}\Delta P~.
\ee
This gives
\ba
\delta \Sigma &=& \frac{1}{c_s^2}\delta P + \left( \frac{1}{c_s^2} \frac{dP}{dr} - \frac{d\Sigma}{dr}\right) \xi_r  = \frac{1}{c_s^2}\delta P + \frac{\Sigma^2 N_r^2}{dP/dr} \frac{i \delta u_r}{\tomega} \label{BV}
\ea
where $\xi_r = i \delta u_r/\tomega$ is the Lagrangian displacement in the $r$ direction, and $N_r$ is the radial Brunt-V\"ais\"ala frequency as given by
\be
N_r^2 = \frac{1}{\Sigma^2}\left(\frac{dP}{dr}\right)^2\left(\frac{d\Sigma}{dP} - \frac{1}{c_s^2}\right)~.
\ee 
Combining the above with the linearized perturbation equations (\ref{perturb1})-(\ref{perturb3}), and eliminating $\delta u_\phi$, we obtain two coupled first-order ODEs appropriate for numerical integration
\begin{subequations}
\ba
\delta h' &=& \left( \frac{\Sigma N_r^2}{P'} + \frac{2m\Omega}{\tomega r}\right) \delta h + \frac{D_s}{\tomega} i \delta u_r \label{foeq1}~,\\
i\delta u_r' &=& \left( \frac{m^2}{\tomega r^2} - \frac{\tomega}{c_s^2}\right) \delta h - \left[\frac{\Sigma N_r^2}{P'} + \frac{m\kappa^2}{2r\Omega\tomega} + \left(\ln r \Sigma\right)' \right] i \delta u_r\label{foeq2}~,
\ea
\end{subequations}
where $\delta h = \delta P/\Sigma$ is the enthalpy perturbation, `` $'$ " denotes $\partial/\partial r$ and 
\be
D_s \equiv \kappa^2 - \tomega^2 + N_r^2~.
\ee
Eliminating $\delta u_r$ we arrive at the second order differential equation for $\delta h$,
\ba
0 &=& \frac{\partial^2}{\partial r^2}\delta h - \frac{d}{d r}\left( \ln \frac{D_s}{r\Sigma}\right) \frac{\partial}{\partial r}\delta h - \left[\frac{m^2}{r^2} + \frac{D_s}{c_s^2} + \frac{2m\Omega}{r\tomega}\frac{d}{d r}\left(\ln \frac{\Omega \Sigma}{D_s} \right)\right]\delta h\nonumber \\
 &\qquad& \qquad~~ -  \left[ \left(\frac{1}{L_S}\right)^2 + \frac{d}{d r}\left(\frac{1}{L_S} \right) -\frac{1}{L_S}\frac{d}{d r}\left(\ln \frac{D_s}{r\Sigma}\right) + \frac{4m\Omega}{\tomega r L_S} - \frac{m^2 N_r^2}{r^2\tomega^2 }\right] \delta h~. \label{secondordereq}
\ea
where 
\be
\frac{1}{L_S} \equiv \frac{\Sigma N_r^2}{dP/dr}~.
\ee It is convenient to eliminate the term proportional to $\delta h'$ in eq. (\ref{secondordereq}) by defining 
\be
A^2 \equiv \frac{D_s}{r\Sigma} \qquad \textrm{and} \qquad  \eta = \frac{\delta h}{A},
\ee
which allows us to rewrite (\ref{secondordereq}) as a wave equation:
\ba
0 &=& \frac{\partial^2}{\partial r^2} \eta - \left[\frac{m^2}{r^2} + \frac{D_s}{c_s^2} + \frac{2m\Omega}{r\tomega}\frac{d}{d r}\left(\ln \frac{\Omega \Sigma}{D_s}\right)  - A \frac{d^2}{dr^2}\frac{1}{A}\right]\eta\nonumber \\
 &\qquad& \qquad~-  \left[ \frac{1}{L_S^2} + \frac{d}{d r}\left( \frac{1}{L_S} \right) -\frac{1}{L_S}\frac{d}{d r}\left(\ln \frac{D_s}{r\Sigma}\right) + \frac{4m\Omega}{\tomega r L_S} + \frac{m^2 N_r^2}{r^2\tomega^2}\right] \eta~, \label{etaeqn}
\ea

Equation (\ref{etaeqn}) forms the basis of our analysis in section 3. If the adiabatic index $\Gamma \equiv \partial \ln P/\partial \ln \Sigma = c_s^2 \Sigma/P$ is constant, one can define the ``entropy", \be
S\equiv P/\Sigma^\Gamma~.\label{Seq}
\ee
Then 
\be
N_r^2 = - \frac{1}{\Gamma \Sigma} \frac{dP}{dr}\frac{d \ln S}{dr}~,
\ee
and 
\be
L_S^{-1} = \frac{1}{\Gamma}\frac{d\ ln S}{dr}~,\label{LSeq}
\ee
and eq.~(\ref{secondordereq}) reduces to eq.~(10) in Lovelace et al (1999). When $N_r\rightarrow 0$ (thus $L_s^{-1} \rightarrow 0$), the terms on the second line of eq.~(\ref{secondordereq}) vanish and we recover the second order perturbation equation for barotropic flows (Goldreich \& Tremaine 1979; Tsang \& Lai 2008).

\section{Reflection of the Corotation Barrier} 

Away from the corotation resonance (where $\tomega = 0$) region, eq.~(\ref{etaeqn}) yields local WKB wave solution $\delta h \propto \exp(i \int k_r dr)$, with $D_s/c_s^2 \simeq - k_r^2$, or
\be
\tomega^2 \simeq \kappa^2 + N_r^2 + k_r^2 c_s^2. 
\ee
Since typically $N_r^2 \lo c_s^2/r^2 \ll \kappa^2 \sim \Omega^2$ (for thin discs), this is the standard dispersion relation for spiral density waves. The inner/outer Lindblad resonances (I/OLR) are defined by $\tomega = \pm \sqrt{\kappa^2 + N_r^2} \simeq \pm \kappa$. Waves can propagate inside the ILR ($r < r_{\rm IL}$) or outside the OLR ($ r > r_{\rm OL}$). between $r_{\rm IL}$ and $r_{\rm OL}$ lies the corotation barrier. 

In this section, we derive the expression for the (complex) reflection coefficient for waves incident upon the corotation barrier and deduce the condition for super-reflection. Our analysis generalizes that given in Tsang \& Lai (2008), which assumed barotropic fluids.

\subsection{Analytical Calculation of the Reflectivity}
Near the corotation resonance $r = r_c$ where $\tomega = 0$, we can rewrite eq.~(\ref{etaeqn}) as
\be
\left[\frac{d^2}{dr^2} - k_{\rm eff}^2 + \frac{2}{q}\left(\frac{d}{dr} \ln \frac{\kappa^2}{\Omega \Sigma} - \frac{2}{L_S}\right) \frac{1}{r-R_c} - \frac{N_r^2}{q^2\Omega^2} \frac{1}{(r-R_c)^2}\right] \eta = 0~,
\ee
where 
$q \equiv -(d\ln\Omega/d\ln r)_c$, $R_c \equiv r_c - i\frac{r_c\omega_i}{q\omega_r}$, and $k_{\rm eff}$ given by
\be
-k_{\rm eff}^2 \equiv \frac{m^2}{r^2} + \frac{D_s}{c_s^2} - A \frac{d^2}{dr^2}\frac{1}{A} + \frac{1}{L_S^2} + \frac{d}{d r}\left( \frac{1}{L_S} \right) -\frac{1}{L_S}\frac{d}{d r}\left(\ln \frac{D_s}{r\Sigma}\right)~,
\ee
is the effective radial wave number without the terms singular at the corotation.
We have introduced a small imaginary part to the wave frequency so that $\omega = \omega_r + i\omega_i$ (with $\omega_i > 0$). Defining 
\be
x \equiv \int_{r_c}^r 2k_{\rm eff} dr ,\qquad \psi \equiv \sqrt{k_{\rm eff}} \eta, \qquad \textrm{and   }\epsilon \equiv \frac{2k_{\rm eff} r_c }{q}\frac{\omega_i}{\omega_r}~,
\ee
we have
\be
\frac{d^2}{dx^2} \psi + \left[-\frac{1}{4} + \frac{\nu}{x + i\epsilon} + \frac{\frac{1}{4} - \mu^2}{(x+i\epsilon)^2}\right]\psi = 0 \label{whittakerdiffeq}~,
\ee
which we recognize as the Whittaker differential equation (Abramowitz \& Stegun 1964). In eq.~(\ref{whittakerdiffeq}) we have defined
\ba
\nu &=&  \left[ \frac{c_s}{q\kappa} \left(\frac{d}{dr} \ln \zeta - \frac{2}{L_S}\right)\right]_c~, \label{nueq} \\
\mu &=&  \frac{1}{2}\left(1 - \frac{4N_r^2}{q^2 \Omega^2} \right)_c^{1/2}~\label{mueq},
\ea
where
\be
\zeta \equiv \frac{\kappa^2}{2\Omega\Sigma}\label{zetaeq}
\ee
is the vortensity for the background flow. When the adiabatic index $\Gamma =$ constant. we can use eq. (\ref{LSeq}) for $L_S^{-1}$, and define the effective vortensity $\zeta_{\rm eff}$ so that
\be
\nu  = \left( \frac{c_s}{q\kappa} \frac{d}{dr}\ln \zeta_{\rm eff} \right)_c~\label{nueq2},
\ee 
with
\be
\zeta_{\rm eff} \equiv \frac{\kappa^2}{2\Omega\Sigma S^{2/\Gamma}}\label{zetaeffeq}~.
\ee
Typically $\kappa \sim \Omega$ (away from the ISCO) and $|N_r| \lo c_s/r \ll \Omega$ (for thin discs), and we have in order of magnitude  $|\nu| \sim c_s/(r\Omega)$ and $|\mu - \tfrac{1}{2}| \sim N_r^2/c_s^2 \lo  c_s^2/(r \Omega)^2$. 

Equation (\ref{whittakerdiffeq}) is solved by the Whittaker functions with indices $\nu$ and $\mu$. The two linearly independent functions of $z = x + i\epsilon$ convenient for construction of connection coefficients are
\be
\psi_- = {\rm W}_{\nu, \mu}(z), \qquad \textrm{and} \qquad \psi_+ = e^{-i\pi\nu} {\rm W}_{-\nu, \mu}(ze^{-i\pi}) + \frac{1}{2} T_0 {\rm W}_{\nu, \mu}(z),
\ee
where $T_0$ is the stokes multiplier (defined below), and $z$ is defined such that ${\rm arg}(z)$ ranges from $0$ to $\pi$. The resulting connection formulae (Tsang \& Lai 2008) are:
\ba
\delta h_- &\sim&
\Biggl\{\begin{array}{ll}
\tfrac{A}{\sqrt{k_{\rm eff}}}\, \exp\left(-\int_{r_c}^{r}\! k_{\rm eff} \,dr \right) 
& \qquad \qquad  ~~\textrm{for } r \gg r_{c}\\
\tfrac{A}{\sqrt{k_{\rm eff}}}\, e^{i\pi\nu}\exp\left(+\int_r^{r_{c}}\! k_{\rm eff} \,dr \right) 
+ \tfrac{A}{\sqrt{k_{\rm eff}}}\, \frac{T_1}{2} e^{-i\pi\nu} \exp\left(-\int_r^{r_{c}}\! k_{\rm eff} \,dr \right) 
&\qquad \qquad~~\textrm{for } r\ll r_{c}~.
\end{array}\label{whitaker1}\\
\delta h_+ &\sim&
\Biggl\{\begin{array}{ll}
\tfrac{A}{\sqrt{k_{\rm eff}}}\, \exp\left(+\int_{r_c}^{r}\! k_{\rm eff} \,dr \right) 
& ~ \textrm{for } r \gg r_{c}\\
\tfrac{A}{\sqrt{k_{\rm eff}}}\, \frac{T_0}{2} e^{i\pi\nu}\exp\left(+\int_r^{r_{c}}\! k_{\rm eff} \,dr \right) +
\tfrac{A}{\sqrt{k_{\rm eff}}}\, \left(1 + \frac{T_1T_0}{4}\right)  e^{-i\pi\nu}\exp\left(-\int_r^{r_{c}}\! k_{\rm eff} \,dr \right) 
&~ \textrm{for } r\ll r_{c}
\end{array}\label{whitaker2}
\ea
where the Stokes multipliers (Heading 1962) are given by
\be 
T_0={2\pi i\over\Gamma(\tfrac{1}{2} - \mu + \nu)\Gamma(\tfrac{1}{2} + \mu+\nu)},\qquad
T_1={2\pi i\,e^{i2\pi\nu}\over\Gamma(\tfrac{1}{2} - \mu -\nu)\Gamma(\tfrac{1}{2} + \mu -\nu)}.\label{stokes}
\ee


The connection formulae (\ref{whitaker1})-(\ref{whitaker2}) here are the same as eqs. (42)-(43)
of Tsang \& Lai (2008), the differences lie in the expressions for $\nu$ [eq. (\ref{nueq})], $T_0$ and $T_1$ [eq. (\ref{stokes})]. For barotropic fluids, $L_S \rightarrow \infty$ and $N_r^2 \rightarrow 0$ (and thus $\mu \rightarrow 1/2$), our expressions reduce to those given in Tsang \& Lai (2008). 
Using the connection formulae (\ref{whitaker1})-(\ref{whitaker2}) and the connection formulae for the Lindblad resonances given by eqs. (32)-(35) in Tsang \& Lai (2008), we can obtain the reflection and transmission coefficients for waves incident upon the corotation barrier for $r < r_{\rm IL}$: 
\ba
&&{\cal R}={ 1+{1\over 4}\,e^{-i2\pi\nu}e^{-2\Theta_{\rm II}}\left(1+{1 \over 4} T_0 T_1\right)
+{i\over 4} T_1 e^{-i2\pi\nu}e^{-2\Theta_{\rm IIa}}
-{i\over 4} T_0e^{-2\Theta_{\rm IIb}}
\over
1-{1\over 4}\,e^{-i2\pi\nu}e^{-2\Theta_{\rm II}}\left(1+{1 \over 4}  T_0 T_1\right)
-{i\over 4} T_1 e^{-i2\pi\nu}e^{-2\Theta_{\rm IIa}}
-{i\over 4} T_0 e^{-2\Theta_{\rm IIb}}}~,\label{fullR}\\
&&{\cal T}={-i\, e^{-\Theta_{\rm II}} e^{i\pi\nu}
\over
1-{1\over 4}\,e^{-i2\pi\nu}e^{-2\Theta_{\rm II}}\left(1+{1 \over 4} T_0 T_1\right)
-{i\over 4} T_1 e^{-i2\pi\nu} e^{-2\Theta_{\rm IIa}}
-{i\over 4}T_0 e^{-2\Theta_{\rm IIb}}}~,\label{fullT}
\ea
where 
\be
\Theta_{\rm II} = \int_{r_{\rm IL}}^{r_{\rm OL}} \sqrt{-k_{\rm eff}^2}\, dr, 
\qquad \Theta_{\rm IIa} = \int_{r_{\rm IL}}^{r_{c}} \sqrt{-k_{\rm eff}^2}\, dr, \textrm{   and}
\qquad \Theta_{\rm IIb} = \int_{r_{c}}^{r_{\rm OL}} \sqrt{-k_{\rm eff}^2}\, dr.
\ee
%

For $|\nu| \ll 1$ and $|\mu - \tfrac{1}{2}| \ll 1$ eq.~(\ref{fullR}) can be simplified to
\be
{\cal R} = { e^{\Theta_{\rm II}} + {1 \over 4} e^{-\Theta_{\rm II}}
\over { e^{\Theta_{\rm II}} - {1 \over 4} e^{-\Theta_{\rm II}}}} + { e^{+2\Theta_{\rm IIb}} + i -{1\over 4}  e^{-2\Theta_{\rm IIb}}  \over \left(e^{\Theta_{\rm II}} - {1 \over 4} e^{-\Theta_{\rm II}}\right)^2} \pi \nu  +   {e^{+2\Theta_{\rm IIb}} -{1\over 4}  e^{-2\Theta_{\rm IIb}}  \over \left(e^{\Theta_{\rm II}} - {1 \over 4} e^{-\Theta_{\rm II}}\right)^2} \pi (\mu - \tfrac{1}{2})+ {\cal O}[\nu^2, (\mu- \tfrac{1}{2})^2] \label{Rapprox}~.\\
\ee
For $\Theta_{\rm IIb} \gg 1$ and $\Theta_{\rm IIa} \gg 1$ this further reduces to 
\be
{\cal R} - 1 \simeq \left(\nu + \mu - \frac{1}{2}\right) \pi e^{-2\Theta_{\rm IIa}}~.\label{Rveryapprox}
\ee
Thus, super-reflection occurs when $\nu + \mu + \tfrac{1}{2}\simeq \nu > 0$ [since $|\mu - \tfrac{1}{2}|$ is much smaller than $|\nu|$; see eqs (\ref{nueq}) - (\ref{mueq})] .

\subsection{Numerical Calculation of Reflectivity}

We can also calculate the reflectivity numerically by integrating eqs.~(\ref{foeq1}) - (\ref{foeq2}). To this end, we assume an outgoing wave at some radius $r_{\rm out} \gg r_{\rm OL}$, motivated by the wave equation (\ref{etaeqn}):
\be
\delta h \propto \frac{A}{\sqrt{k_r}} \exp \left[i\int_{r_{\rm OL}}^{r} k_rdr + i\frac{\pi}{4}\right]~.
\ee
where $k_r$ is the full radial wave-number given by 
\be
-k_r^2\equiv \frac{m^2}{r^2} + \frac{D_s}{c_s^2} + \frac{2m\Omega}{r\tomega}\frac{d}{d r}\left(\ln \frac{\Omega \Sigma}{D_s}\right)  - A \frac{d^2}{dr^2}\frac{1}{A} + \frac{1}{L_S^2} + \frac{d}{d r}\left( \frac{1}{L_S} \right) -\frac{1}{L_S}\frac{d}{d r}\left(\ln \frac{D_s}{r\Sigma}\right) + \frac{4m\Omega}{\tomega r L_S} + \frac{m^2 N_r^2}{r^2\tomega^2}~.
\ee
This gives the outer boundary condition at $r = r_{\rm out}$
\be
\delta h'(r_{\rm out}) = \left(i k_r + \frac{d}{dr} \ln A - \frac{1}{2}\frac{d}{dr} \ln k_r\right) \delta h(r_{\rm out})~.\label{outbc}
\ee
At some inner radius $r_{\rm in} \ll r_{\rm IL}$, the solution takes the form 
\be
\delta h\propto \frac{A}{\sqrt{k_r}}\left( \exp\left[ i\int_{r}^{r_{\rm IL}} k_r\,dr + i\frac{\pi}{4}\right] + {\cal R} \exp\left[ -i\int_{r}^{r_{\rm IL}} k_r\,dr - i\frac{\pi}{4}\right]\right)~,
\ee
and the reflection coefficient can be obtained from
\be
|{\cal R}| = \left| \frac{\left(\tfrac{d}{dr}\ln A - \tfrac{1}{2}\tfrac{d}{dr}\ln k_r - i k_r \right)\delta h - \delta h'}{\left(\tfrac{d}{dr}\ln A - \tfrac{1}{2}\tfrac{d}{dr}\ln k_r + i k_r \right)\delta h - \delta h'}\right|_{r_{\rm in}}~.
\ee

\begin{figure}
\centering
\begin{tabular}{ccc}
\epsfig{file=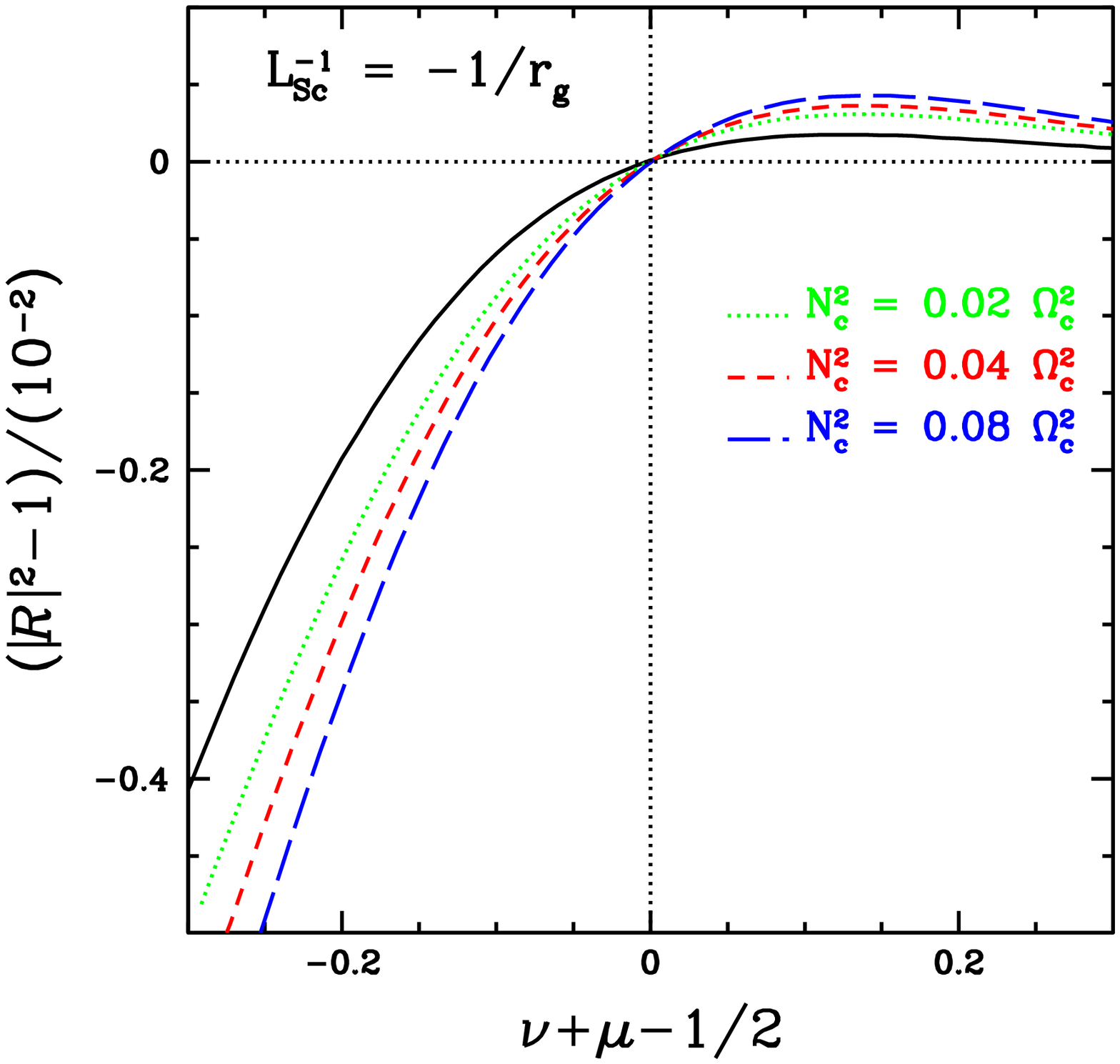,width=0.5\linewidth,clip=} &
\epsfig{file=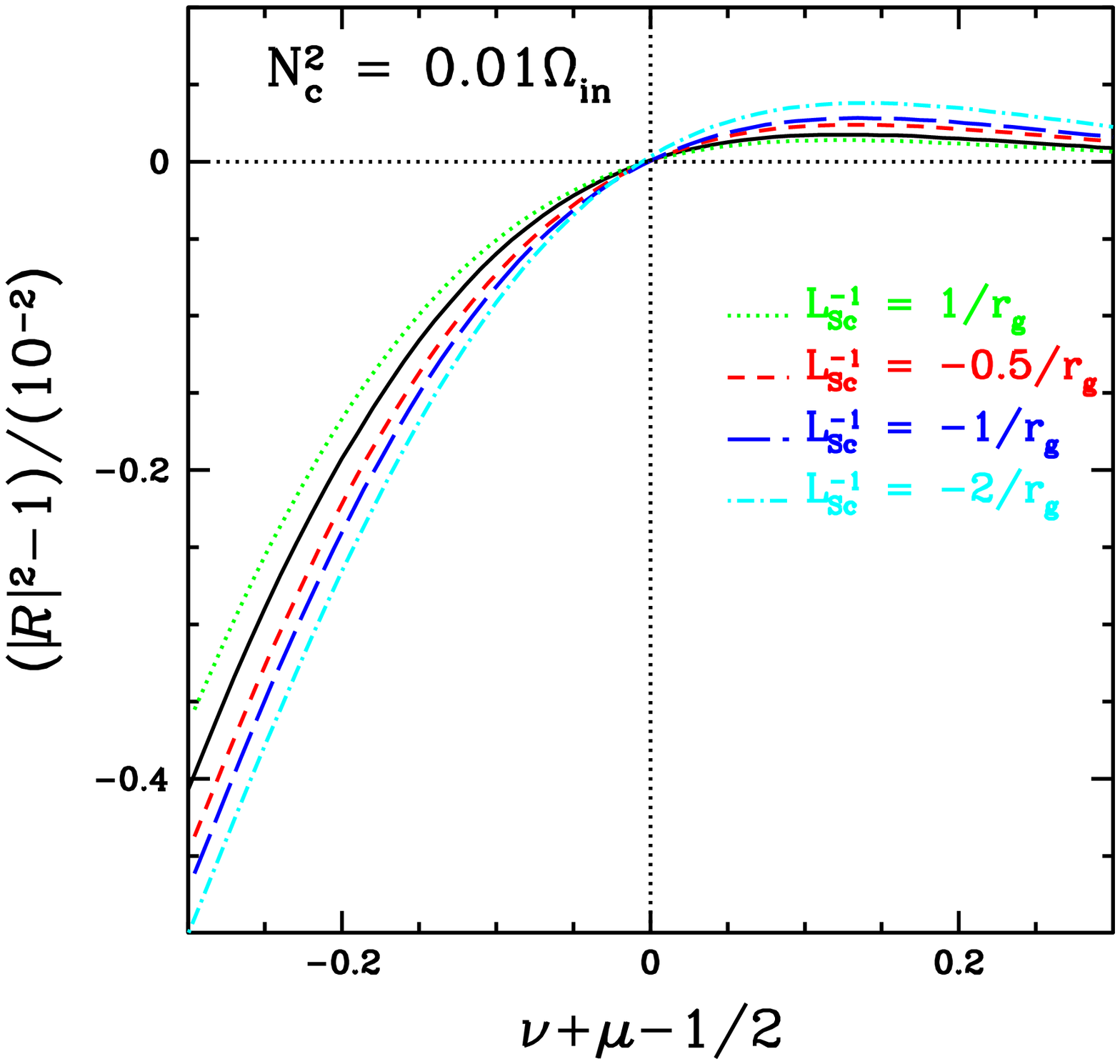,width=0.5\linewidth,clip=}
\end{tabular}
\caption{Reflectivity of the corotation barrier in a Keplerian disc. The disc background profiles are  given by $\Omega \propto r^{-3/2}$, $L_{S} = L_{Sc} (r/r_c)^{-1}$, $N_r^2 = N_c^2 (r/r_c)^{-3/2}$, and the sound speed given by $c_s/(r_c \Omega_c) = 0.1$. The horizontal axis gives $\nu + \mu - \tfrac{1}{2}$ [see eqs.~(\ref{Rapprox}) - (\ref{Rveryapprox})]. The parameter $\nu$ is varied by changing the density profile $\Sigma \propto (r/r_c)^{-p}$. The reflectivity for barotropic fluids (Tsang \& Lai 2008) is recovered for $N_c^2 = 0$ and $L_{Sc}^{-1} = 0$, shown as the solid line.}
\label{Rfig}
\end{figure}

Fig.~\ref{Rfig} gives some examples of the reflectivity for a simple power-law disc model. We parameterize the relevant disc profiles by assuming $\Sigma \propto r^{-p}$, $N_r^2 = N_c^2 (r/r_c)^{-b}$ and $L_S = L_{Sc}(r/r_c)^{-\gamma}$, and we fix the sound speed to $c_s/(r\Omega) = 0.1$. Note that for non-barotropic fluids $P \neq \Sigma c_s^2$, thus we allow $N_r^2$ and $L_S^{-1} =\Sigma N_r^2/P'$ to vary independently. In the examples depicted in Fig.~\ref{Rfig}, we fix $N_r(r)$ and $L_{S}(r)$, but vary the density index, $p$, to change the critical parameter $\nu$ [see eq. (\ref{nueq})]. In general, as seen from eqs.~(\ref{foeq1}) - (\ref{foeq2}) or eq.~(\ref{etaeqn}), the result depends on $N_c$, $L_{Sc}$, $b$, and $\gamma$, but we find that the dependence on $b$ and $\gamma$ to be rather weak. In agreement with the analytic expression in Section 3.1 [eq.~(\ref{Rveryapprox})], we find that the super-reflection ($|{\cal R}| > 1$) is achieved when $\nu + \mu - \tfrac{1}{2} \simeq \nu > 0$.

\section{Calculation of Global Overstable P-Modes in Black Hole Accretion Discs}

The result of Section 3 shows that when $\nu + \mu - \tfrac{1}{2} > 0$, waves impinging upon the corotation barrier are super-reflected. Supposing there exists a reflecting boundary at the inner disc radius $r_{\rm in} = r_{\rm ISCO}$, normal p-modes can be produced, with waves trapped between $r_{\rm in}$ and $r_{\rm IL}$. In the WKB approximation the mode growth rate $\omega_i$ is directly related to the reflectivity $|{\cal R}|$ (Tsang \& Lai 2008)
\be
\omega_i \simeq \left( \frac{|{\cal R}| - 1}{|{\cal R}| + 1}\right)\left[\int_{r_{\rm in}}^{r_{\rm IL}} \frac{|\tomega_r|}{c_s \sqrt{\tomega_r^2 - \kappa^2}}dr\right]^{-1}
\ee
where $\tomega_r = {\rm Re}(\omega) - m\Omega$. More accurate calculation of the p-mode frequency $\omega = \omega_r + i\omega_i$ requires solving the complex eigenvalue problem based on eqs.~(\ref{foeq1}) - (\ref{foeq2}).

\subsection{Background Disk Structure}

As an illustration of the global p-mode calculation, we consider the standard $\alpha$-disc model. For the inner region of the disc we are most concerned with, radiation pressure dominates gas pressure, and the opacity is primarily due to electron scattering.  However it is well known that with  the standard viscosity prescription for the viscous stress tensor $\sigma_{r\phi} = -\alpha P_{\rm total}$, this inner disc solution is thermally unstable (Shakura \& Sunyaev 1976). We therefore also consider a slightly modified disk model where $\sigma_{r\phi} = -\alpha P_{\rm gas}$ is adopted --  this disc solution is thermally stable in the inner region (Lightman 1974). Consistent with our perturbation analysis, we use the Paczynski-Wiita potential $\Phi = - GM/(r-2r_g)$ (where $r_g = GM/c^2$) to mimic the general relativistic effect (Pacyznski \& Wiita 1980). 

\begin{figure}
\centering
\includegraphics[width=13cm]{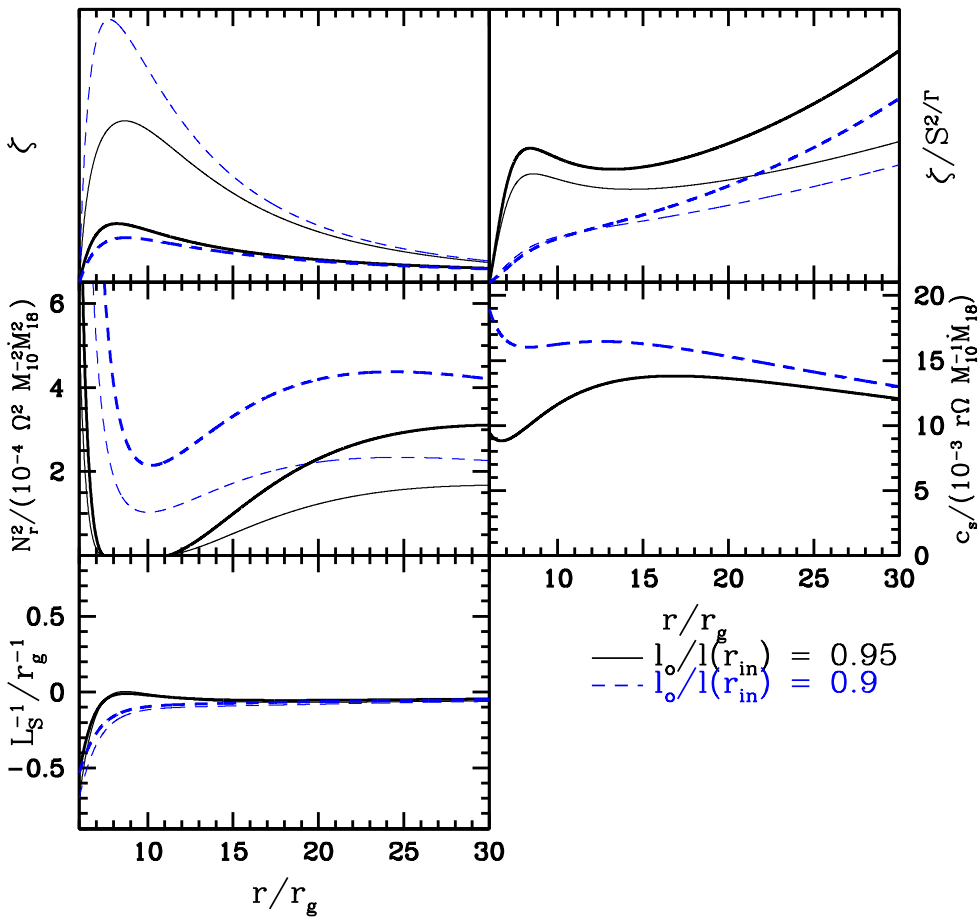}
\caption{
The background disc solutions for $\alpha$-disc models. The depicted profiles are: the vortensity $\zeta = \kappa^2/(2\Sigma\Omega)$ and the modified vortensity $\zeta_{\rm eff} = \zeta/S^{2/\Gamma}$ [both in arbitrary units; see eqs.~(\ref{zetaeq}) and (\ref{zetaeffeq})]; the squared radial Brunt-V\"ais\"ala frequency $N_r^2$, the inverse entropy length-scale $L_S^{-1}$ [see eqs.(\ref{Seq}) - (\ref{LSeq})], and the sound speed $c_s$. The adiabatic index is assumed to be $\Gamma = 1.4$. The solid and dashed lines denote different angular momentum eigenvalues $l_o$. The thick lines show the profiles for the $\sigma_{r\phi} = -\alpha P_{\rm gas}$ prescription, while the thin lines show the profiles for the $\sigma_{r\phi} = -\alpha P_{\rm total}$ prescription.
}
\label{alphadiskfig}
\end{figure}

With the viscosity prescription $\sigma_{r\phi} = - \alpha P_{\rm total}$ the relevant background disk profiles are 
\begin{subequations}
\ba
\Sigma(r) 
&=& (6.95 \textrm{ g  cm}^{-2}) M_{10} \dot{M}_{18}^{-1} \alpha^{-1} \left(\tfrac{r}{r_g}\right)^{3/2} \left(1 - \tfrac{2r_g}{r}\right)^3 \left(1-\tfrac{2r_g}{3r}\right)^{-2}{\cal J}^{-1}\\  
P(r)
 &=& (3.24\times 10^{21} \textrm{ erg cm}^{-2}) M_{10}^{-1} \dot{M}_{18} \alpha^{-1} \left( \tfrac{r}{r_g}\right)^{-3/2}\left(1 - \tfrac{2r_g}{r}\right)^{-1} \cal{J}\\
H(r) 
&=& (1.06\times 10^6 \textrm{ cm}) \dot{M}_{18} \left(1 - \tfrac{2r_g}{r}\right)^{-1} \left(1-\tfrac{2r_g}{3r}\right)\cal{J}~.
\ea
\end{subequations}
where  $M_{10} = M/(10 M_\odot)$, $\dot{M} = (10^{18} g s^{-1}) \dot{M}_{18}$ is the accretion rate, $H(r)$ is the vertical scale height, and ${\cal J} \equiv 1 - l_o/l(r)$ with $l(r) = \Omega r^2$. The constant $l_o$ specifies the specific angular momentum absorbed at the inner edge of the disc per unit accreting mass; one typically expects $l_o \leq l(r_{\rm in})$ [$l_o = l(r_{\rm in})$ is the so-called zero-torque condition]. 

The viscosity prescription with $\sigma_{r\phi} = -\alpha P_{\rm gas}$ yields
\begin{subequations}
\ba
\Sigma(r) 
&=& (7.02\times10^4 \textrm{ g  cm}^{-2}) M_{10}^{-\tfrac{2}{5}}\dot{M}_{18}^{\tfrac{3}{5}} \alpha^{-\tfrac{4}{5}} \left(\tfrac{r}{r_g}\right)^{-\tfrac{3}{5}} \left(1-\tfrac{2r_g}{r}\right)^{-\tfrac{1}{5}}\left(1 - \tfrac{2r_g}{3r}\right)^{-\tfrac{1}{5}} {\cal J}^{\tfrac{3}{5}}~, \\ 
P(r) 
 &=& (3.27\times 10^{25} \textrm{ erg cm}^{-2}) M_{10}^{-\tfrac{12}{5}}\dot{M}_{18}^{\tfrac{13}{5}} \alpha^{-\tfrac{4}{5}}\left(\tfrac{r}{r_g}\right)^{-\tfrac{18}{5}}\left(1-\tfrac{2r_g}{r}\right)^{-\tfrac{21}{5}}\left(1-\tfrac{2r_g}{3r}\right)^{\tfrac{9}{5}}{\cal J}^{\tfrac{13}{5}}~,\\
H(r) 
&=& (1.06\times 10^6 \textrm{ cm}) \dot{M}_{18} \left(1 - \tfrac{2r_g}{r}\right)^{-1} \left(1-\tfrac{2r_g}{3r}\right)\cal{J}~.
\ea
\end{subequations}

In both cases we obtain the 2-dimensional adiabatic sound speed by $c_s^2 = \Gamma P/\Sigma$, giving 
\be
\frac{c_s}{r\Omega} = 0.72 \Gamma^{1/2} M_{10}^{-1} \dot{M}_{18} \left(\tfrac{r}{r_g}\right)^{-1} \left(1 - \tfrac{2r_g}{3r} \right) \left( 1- \tfrac{2r_g}{r}\right)^{-1}{\cal J}~.\label{soundspeedeq}
\ee
For simplicity, we adopt $\Gamma = 1.4$ in our calculations (using somewhat different values do not affect our results in Section 4.2). 
Fig.~\ref{alphadiskfig} depicts the disc background profiles important for our p-mode calculations.

\subsection{Growing Eigenmodes}

In addition to the outgoing boundary condition (\ref{outbc}) at some $r_{\rm out} \gg r_{\rm OL}$, it is necessary to impose an appropriate inner boundary condition  (at $r_{\rm in} = r_{\rm ISCO}$) in order to calculate the global p-modes trapped between $r_{\rm in}$ and $r_{\rm IL}$. Unfortunately this inner boundary condition is uncertain: the large radial velocity of the transonic flow around $r_{\rm ISCO}$ leads to energy loss of the wave, while the sharp density gradient at $r_{\rm ISCO}$ provides a partially reflecting inner boundary (see Lai \& Tsang 2009); in real black-hole accretion flows, a large magnetic flux accumulation inside $r_{\rm ISCO}$ can make the inner disc edge an even better reflector for waves. Here, to focus on the role of the corotational instability, we adopt the free boundary condition (zero Lagrangian pressure perturbation) at $r_{\rm in} = r_{\rm ISCO}$, i.e.
\be
\Delta P = \left(\delta P + \frac{dP}{dr} \frac{i\delta u_r}{\tomega}\right)_{r_{\rm in}} = 0~.\label{inbc}
\ee

\begin{figure}
\centering
\begin{tabular}{ccc}
\epsfig{file=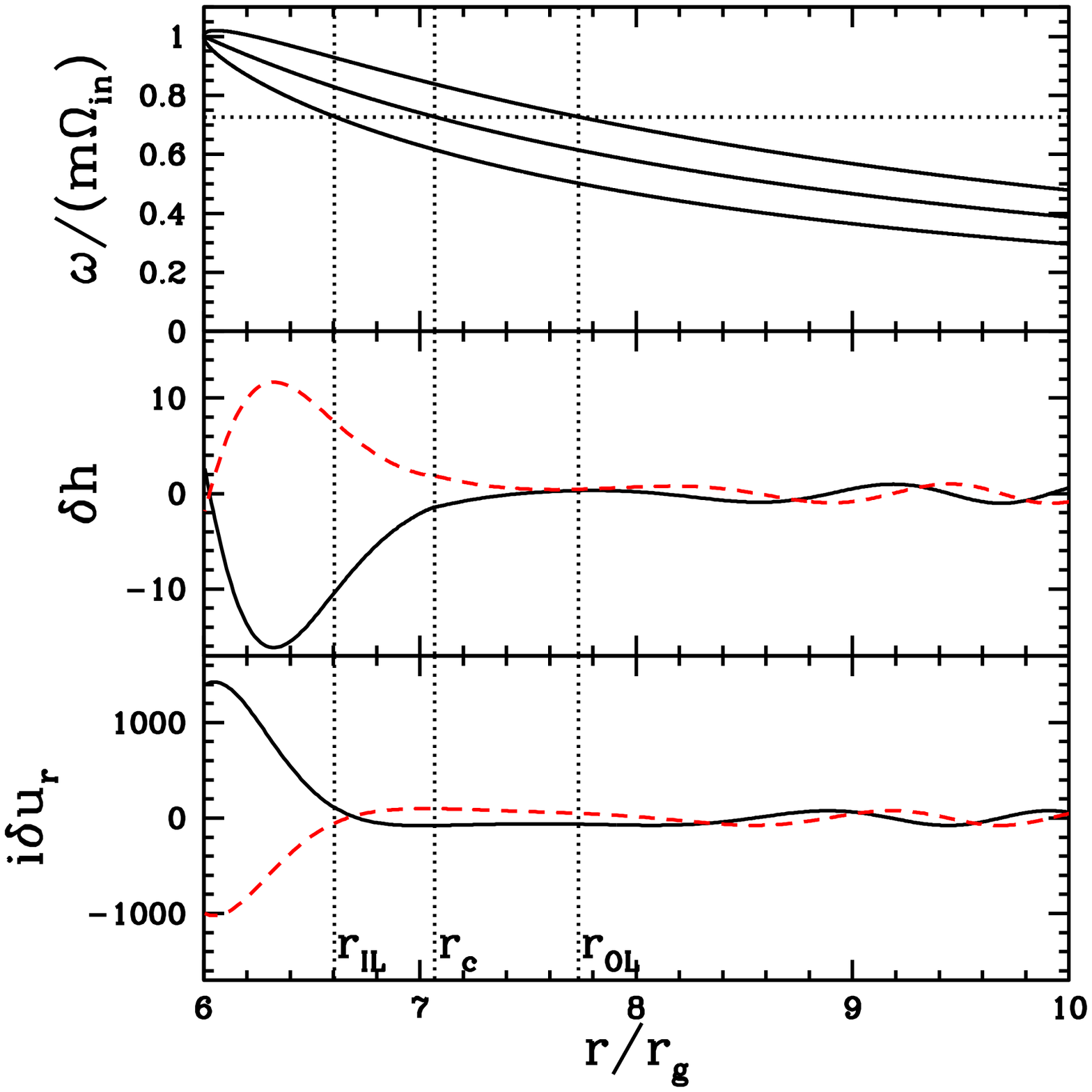,width=0.5\linewidth,clip=} &
\epsfig{file=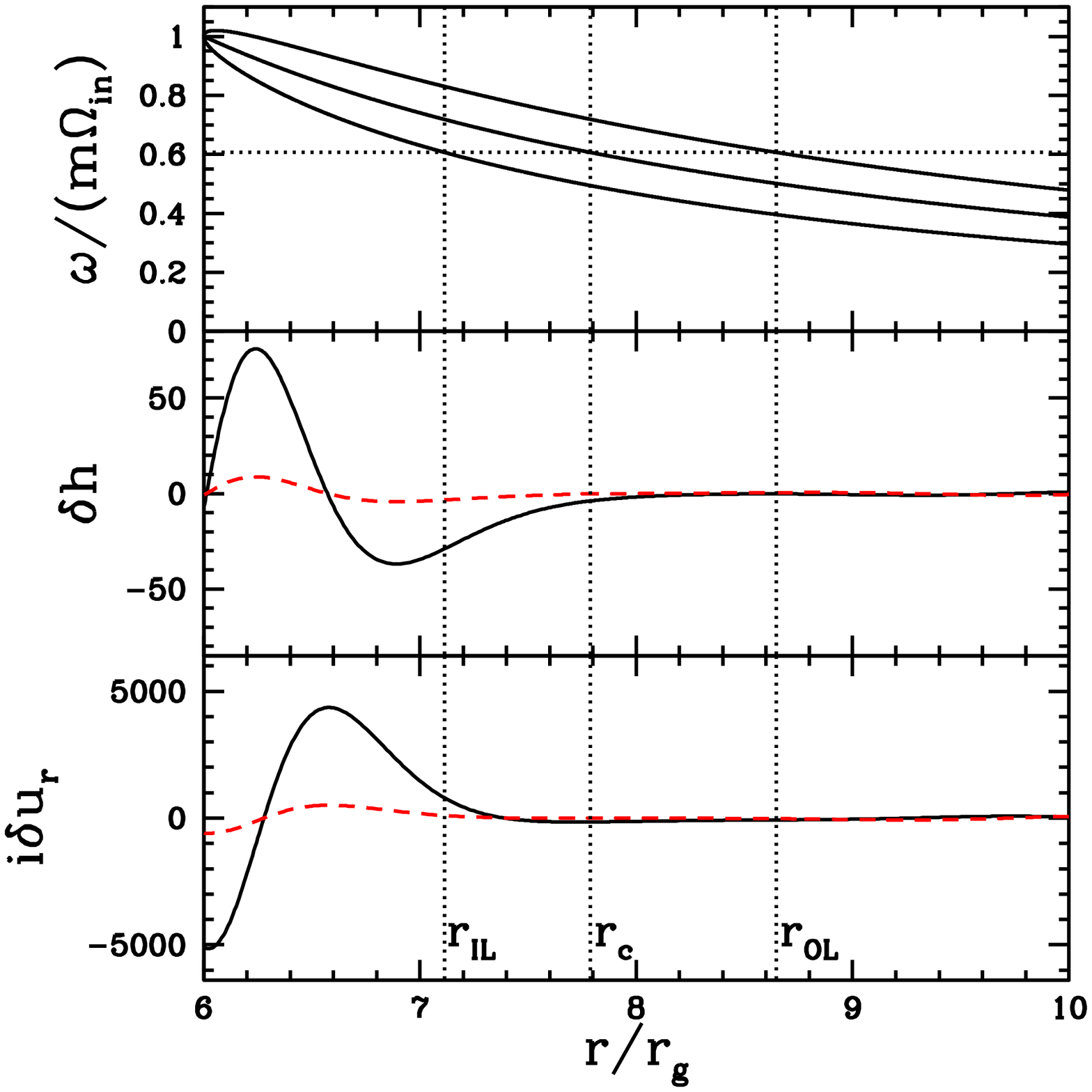,width=0.5\linewidth,clip=}
\end{tabular}
\caption{Eigenunctions of disc p-modes with $m = 3$. The disc model parameters are $\dot{M}_{18} = 3$, and $l_o/l(r_{\rm in}) = 0.95$, with the viscous stress tensor given by $\sigma_{r \phi} = -\alpha P_{\rm gas}$. The left panel shows the propagation diagram and eigenfunctions for the zero-node (in the trapping region) mode, with eigenfrequency of  $\omega/\Omega_{\rm in} = 2.18 + i3.0\times 10^{-3}$. The right plot shows the propagation diagram and eigenfunctions the single-node mode with $\omega/\Omega_{\rm in} = 1.82 + i2.3\times 10^{-4}$. For the eigenfunctions, the solid lines denote the real parts, while the dashed lines denote the imaginary parts. }
\label{wavefunfig1}
\end{figure}

With eq. (\ref{outbc}) and eq. (\ref{inbc}) we employ the standard shooting method (Press et. al 1995) with eqs.~(\ref{foeq1}) and (\ref{foeq2}), to solve for the eigenvalue $\omega = \omega_r + i \omega_i$. Table 1 summarizes the results for different background disc parameters, and example wavefunctions for the zero and one-node eigenmodes are shown in Fig.~\ref{wavefunfig1}. We see that the trapping region extends from the inner edge of the disc at $r_{\rm ISCO}$ to the inner Lindblad resonance $r_{\rm IL}$. The wave is evanescent in the corotation barrier region between $r_{\rm IL}$ and $r_{\rm OL}$, and tunnels out to the propagation region ($r > r_{\rm OL}$).
In the following we will focus on the 0-node modes since they 
have growth rates much larger than the 1-node modes.

\begin{figure}
\centering \includegraphics[width=13cm]{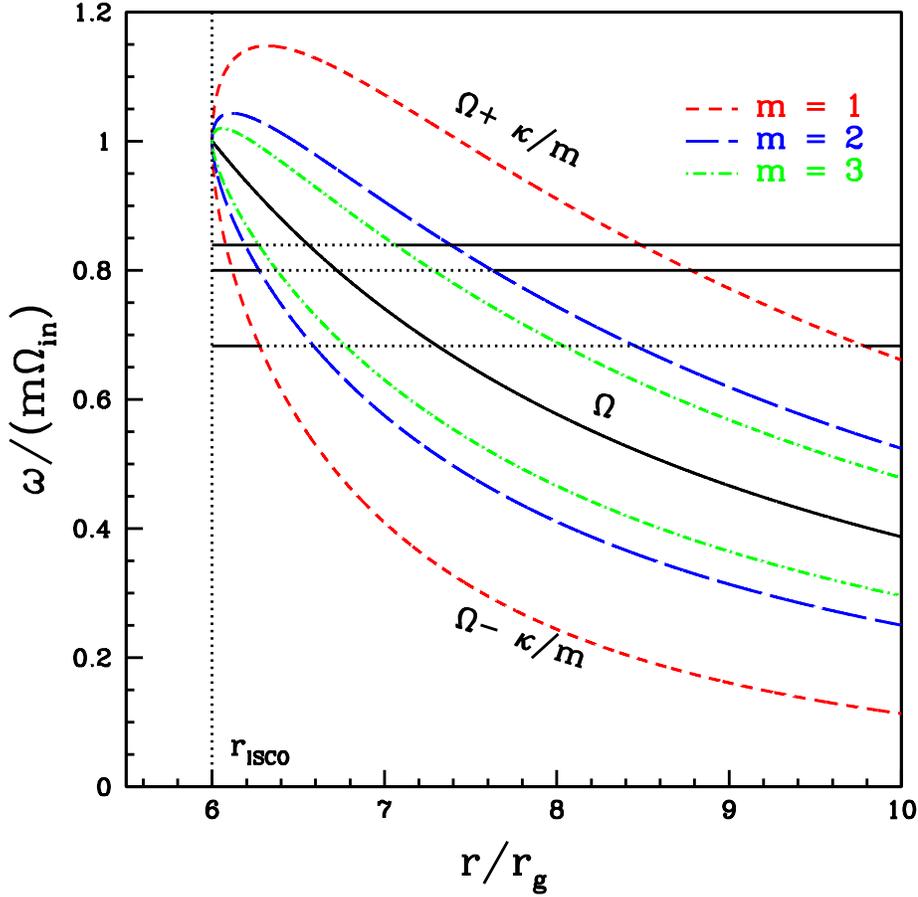}
\caption{The propagation diagram for disc p-modes with various values
  of $m$. The solid curve shows the disc rotation profile $\Omega$,
  while the various dashed curves show $\Omega+\kappa/m$ (above the
  $\Omega$ curve) and $\Omega-\kappa/m$ (below the $\Omega$ curve).
  The three horizontal lines show the representative values of the
  mode frequency [in units of $m\Omega_{\rm in}$, where $\Omega_{\rm
      in}=\Omega(r_{\rm in})$] for $m=1,2,3$ (from bottom to top).
  The corotation resonance is determined by $\omega/m = \Omega$, the
  inner Lindblad resonance by $\omega/m = \Omega - \kappa/m$ and the
  outer Lindblad resonance by $\omega/m = \Omega + \kappa/m$.  A
  mode is trapped between $r_{\rm in}=r_{\rm ISCO}$ and $r_{\rm IL}$,
  and is evanescent between $r_{\rm IL}$ and $r_{\rm OL}$ (the
  horizontal dotted lines).
}
\label{propfig}
\end{figure}

From Table 1 we see that for a given disc model, the (real) mode
pattern freqency $\omega_r/m$ increases only slightly as $m$
increases, while the growth rate more rapidly increases with increasing
$m$.  In particular, the $m=1$ mode has a much smaller growth rate
than the higher-$m$ modes. These features can be easily understood by
examining the propagation diagram (Fig.~\ref{propfig}). For small $m$,
the wave trapping region between $r_{\rm ISCO}$ and $r_{\rm IL}$ is
slimmer, thus to ``contain'' the same number of wavelengths in the
trapping region, the pattern frequency must be lower. On the other
hand, the wider corotation barrier for small $m$ implies that
only a small amount of wave energy can tunnel through the barrier, giving
rise to smaller corotational wave absorption and a slow mode growth
rate. As can be seen from Fig.~\ref{propfig}, the difference in the
width of the evanescent regions is greatest between the $m=1$ and
$m=2$; the lower pattern frequency of lower $m$
modes also helps to widen the evanescent region. These explain why the
$m=1$ mode has such a small growth rate compared to the other modes.

Table 1 shows that for the $\alpha$-disc models considered, 
the mode frequency decreases slightly as $\dot M$ increases (while
keeping the other disc parameters fixed). This results from 
the increase of the disc sound speed $c_s$ [see eq.~(\ref{soundspeedeq})].
Table 1 also shows that the $m=2$ and $m=3$ modes in each disc model
have roughly 2:3 commensurate frequencies, ranging from
$\omega_{m=3}/\omega_{m=2} \simeq 1.57$ to $1.69$ for the disc background
models considered. This has implications for the observations of 
high-frequency QPOs (see section 5).

\begin{figure}
\centering
\includegraphics[width=13cm]{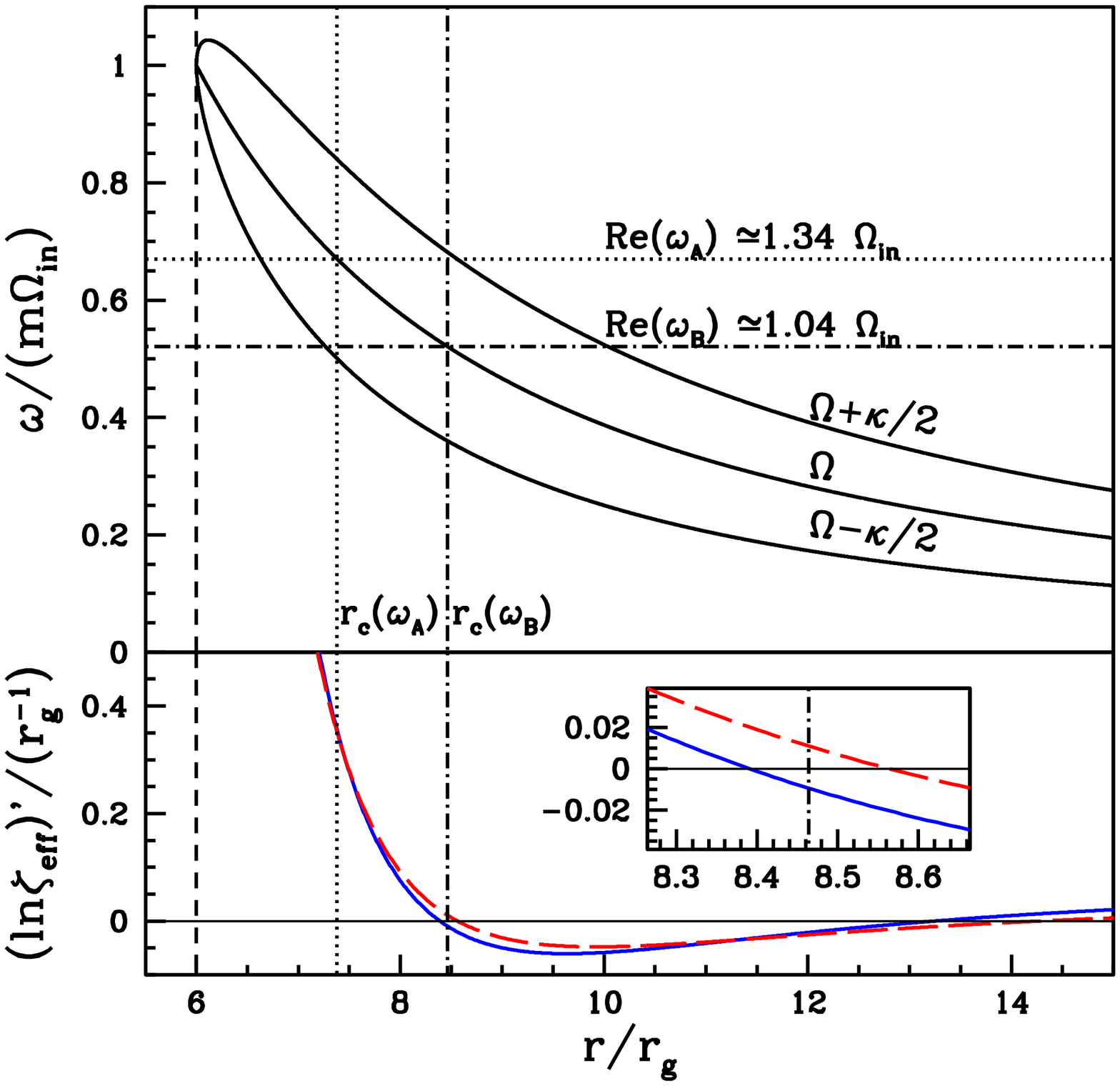}
\caption{The propagation diagram for $m=2$ disc p-modes 
and the derivative of the effective disc vortensity profiles. 
In the lower panel, the $\zeta_{\rm eff}$ profiles
are shown for disc models with the viscosity prescription
$\sigma_{r\phi}=-\alpha P_{\rm gas}$ (solid curve) and
$\sigma_{r\phi}=-\alpha P_{\rm total}$ (dashed curve), and the other 
disc parameters are $\dot M_{18}=3$, $l_o/l(r_{\rm in})=0.95$.
In the upper panel, the two horizontal lines
give the real eigenfrequencies of the 0-node mode,
${\rm Re}(\omega) \simeq 1.34\Omega_{\rm in}$ (for both disc models) 
and the 1-node mode,
${\rm Re}(\omega) \simeq 1.04\Omega_{\rm in}$ (for the 
$\sigma_{r\phi}=-\alpha P_{\rm total}$ disc model only).
Note that growing modes can be found only if 
$d\ln\zeta_{\rm eff}/dr > 0$ at the corotation, thus the 
${\rm Re}(\omega) \simeq 1.44\Omega_{\rm in}$ mode exists
for both disc models, while the 
${\rm Re}(\omega) \simeq 1.04\Omega_{\rm in}$ mode exists 
only in one of the disc models.
The inset of the lower panel shows a magnified view of the
derivative of the effective vortensities at the corotation point of
the ${\rm Re}(\omega) \simeq 1.04\Omega_{in}$ mode. }
\label{vortfig}
\end{figure}

Fig.~\ref{vortfig} shows the propagation diagram of $m=2$ modes and
the effective vortensity gradient profiles for both viscosity-law
($\sigma_{r\phi}=-\alpha P_{\rm total}$ vs $\sigma_{r\phi}=-\alpha P_{\rm gas}$)
background disc models, with $\dot{M}_{18} = 3$, $l_o/l(r_{\rm in}) = 0.95$ and
$\Gamma = 1.4$.  The zero-node modes for both models occur at ${\rm
  Re}(\omega) \simeq 1.34\Omega_{\rm in}$, which has a positive
effective vortensity slope at the corotation radius. However only the
disc model with the $\sigma_{r\phi}=-\alpha P_{\rm total}$ prescription has a
growing 1-node mode at ${\rm Re}(\omega) \simeq 1.04\Omega_{\rm in}$
(see Table 1). For the disc model with the $\sigma_{r\phi}=-\alpha P_{\rm
  gas}$ prescription, such a (real) frequency would give negative
effective vortensity gradient at $r_c$, thus the corotation acts to
damp the mode (see the inset of Fig.~\ref{vortfig}).

Overall, our numerical calculation of the global disc p-modes 
is in agreement with our analysis given in Section 3, i.e.,  
wave absorption at the corotation resonance gives rise to growing
p-modes when the gradient of the effective vortensity is positive
at corotation.

%
\begin{table*}
\centering
\begin{minipage}{140mm}
\caption{Overstable p-mode frequencies for various disc models.}
\begin{tabular}{| c  c || r@{ + $i$}l | r@{ + $i$}l | r@{ + $i$}l | r@{ + $i$}l |}
\hline\hline
\multicolumn{10}{|c|}{Mode Eigenfrequencies ($\omega/\Omega_{\rm in}$) for $\sigma_{r \phi} = - \alpha P_{\rm total}$}\\
\hline
\multicolumn{2}{|c||}{mode} 
& \multicolumn{2}{|c|}{$\beta = 0.9$, $\dot{M}_{18} = 1$} 
& \multicolumn{2}{|c|} {$\beta = 0.9$, $\dot{M}_{18} = 3$} 
&  \multicolumn{2}{|c|} {$\beta = 0.95$, $\dot{M}_{18} = 1$}  
&  \multicolumn{2}{|c|} {$\beta = 0.95$, $\dot{M}_{18} = 3$} \\ \hline\hline
\multirow{2}{*}{$m = 1$}
& 0-node 
& $0.59$ & $2.2\times10^{-7}$ 
&$0.39$ & $4.0\times10^{-6}$ 
& $0.68$ & $2.2\times 10^{-8}$ 
& $0.52$ & $1.4\times10^{-7}$ \\
& 1-node 
& $0.37$ & $2.4\times10^{-9}$ 
&$0.17$ & $3.1\times10^{-9}$ 
& \multicolumn{2}{|c|}{--}
& \multicolumn{2}{|c|}{--}   \\ \hline
\multirow{2}{*}{$m = 2$}
& 0-node 
& $1.46$ & $4.0\times10^{-4}$ 
&$1.13$ & $1.3\times10^{-3}$ 
& $1.60$ & $1.2\times 10^{-4}$ 
& $1.35$ & $6.2\times10^{-4}$ \\
& 1-node 
& $1.18$ & $4.2\times10^{-6}$ 
&$0.80$ & $5.6\times10^{-5}$ 
& $1.37$ & $5.5\times 10^{-7}$ 
& $1.04$ & $2.5\times10^{-6}$ \\ \hline
\multirow{2}{*}{$m = 3$}
& 0-node 
& $2.34$ & $2.6\times10^{-3}$ 
&$1.91$ & $5.5\times10^{-3}$ 
& $2.52$ & $1.2\times 10^{-3}$ 
& $2.20$ & $3.4\times10^{-3}$ \\
& 1-node 
& $2.01$ & $1.8\times10^{-4}$ 
&$1.48$ & $7.7\times10^{-4}$ 
& $2.25$ & $4.4\times 10^{-5}$ 
& $1.83$ & $2.6\times10^{-4}$ \\ \hline
\multirow{2}{*}{$m = 4$}
& 0-node 
& $3.23$ & $6.0\times10^{-3}$ 
&$2.71$ & $1.0\times10^{-2}$ 
& $3.44$ & $3.2\times 10^{-3}$ 
& $3.06$ & $7.3\times10^{-3}$ \\
& 1-node 
& $2.86$ & $9.0\times10^{-4}$ 
&$2.20$ & $2.4\times10^{-3}$ 
& $3.15$ & $3.2\times 10^{-4}$ 
& $2.63$ & $1.2\times10^{-3}$ \\ \hline \hline
\multicolumn{10}{|c|}{Mode Eigenfrequencies ($\omega/\Omega_{\rm in}$) for $\sigma_{r \phi} = -\alpha P_{\rm gas}$}\\
\hline
\multicolumn{2}{|c||}{mode} 
& \multicolumn{2}{|c|}{$\beta = 0.9$, $\dot{M}_{18} = 1$} 
& \multicolumn{2}{|c|} {$\beta = 0.9$, $\dot{M}_{18} = 3$} 
&  \multicolumn{2}{|c|} {$\beta = 0.95$, $\dot{M}_{18} = 1$}  
&  \multicolumn{2}{|c|} {$\beta = 0.95$, $\dot{M}_{18} = 3$} \\ \hline\hline
\multirow{2}{*}{$m = 1$}
& 0-node 
& $0.58$ & $1.1\times10^{-7}$ 
&$0.37$ & $2.3\times10^{-6}$ 
& $0.67$ & $2.3\times 10^{-8}$ 
& \multicolumn{2}{|c|}{--}\\
& 1-node 
& $0.37$ & $9.6\times10^{-10}$ 
& $0.16$ & $1.1\times10^{-9}$ 
& \multicolumn{2}{|c|}{--}
& \multicolumn{2}{|c|}{--}   \\ \hline
\multirow{2}{*}{$m = 2$}
& 0-node 
& $1.44$ & $3.6\times10^{-4}$ 
& $1.11$ & $1.1\times10^{-3}$ 
& $1.59$ & $1.0\times 10^{-4}$ 
& $1.33$ & $4.9\times10^{-4}$ \\
& 1-node 
& $1.18$ & $4.2\times10^{-6}$ 
& $0.79$ & $6.0\times10^{-5}$ 
& $1.36$ & $4.3\times 10^{-7}$ 
& \multicolumn{2}{|c|}{--}\\ \hline
\multirow{2}{*}{$m = 3$}
& 0-node 
& $2.32$ & $2.4\times10^{-3}$ 
& $1.88$ & $5.1\times10^{-3}$ 
& $2.51$ & $1.1\times 10^{-3}$ 
& $2.18$ & $3.0\times10^{-3}$ \\
& 1-node 
& $2.00$ & $1.9\times10^{-4}$ 
& $1.47$ & $8.1\times10^{-4}$ 
& $2.25$ & $4.4\times 10^{-5}$ 
& $1.82$ & $2.3\times10^{-4}$ \\ \hline
\multirow{2}{*}{$m = 4$}
& 0-node 
& $3.21$ & $5.7\times10^{-3}$ 
& $2.68$ & $1.0\times10^{-2}$ 
& $3.44$ & $3.1\times 10^{-3}$ 
& $3.04$ & $6.8\times10^{-3}$ \\
& 1-node 
& $2.85$ & $9.3\times10^{-4}$ 
& $2.19$ & $2.5\times10^{-3}$ 
& $3.14$ & $3.2\times 10^{-4}$ 
& $2.62$ & $1.2\times10^{-3}$ \\ \hline\hline
\end{tabular}
\\A dash indicates that no growing eigenmode could be found, and that the mode is damped. Here the parameter $\beta \equiv l_o/l(r_{\rm in})$ determines the inner torque condition for the background disc. The mode frequencies are independent of the value of the viscosity parameter, $\alpha$, used. 
\end{minipage}
\end{table*}

\section{Discussion}

We have studied the effect of corotation resonance on the adiabatic
diskoseismic p-modes (inertial-acoustic oscillations) of
non-barotropic accretion flows around black holes. 
Our WKB analysis of the reflectivity of the corotation barrier (Section 3),
as well as our numerical calculation of the global disc p-modes (Section 4), show that 
the corotational wave absorption can be significantly modified by the 
non-barotropic effect. In particular, we have showed that super-reflection is
achieved when [see eq.~(\ref{Rveryapprox}) and Fig.~\ref{Rfig}]
\be
\nu+\mu-{1\over 2}>0,
\ee
where $\nu,~\mu$ are defined by eqs.~(\ref{nueq})-(\ref{mueq}). For thin discs, $|\mu-1/2|\ll 1$
and this condition is simply $\nu>0$, or
\be
{d\over dr}\ln\zeta-{2\Sigma N_r^2\over dP/dr}={d\over dr}\ln
\left({\zeta\over S^{2/\Gamma}}\right)>0,
\ee
where $\zeta=\kappa^2/(2\Omega\Sigma)$ is the disc vortesnity,
$N_r$ is the radial Brunt-V\"as\"al\"a frequency, 
and the first equality holds only when the adiabatic index 
$\Gamma=$~constant (in which case $S=P/\Sigma^\Gamma$).
Thus, in the presence of a reflecting (or partially reflecting) 
boundary at the disc inner edge ($r=r_{\rm in}$),
the non-axisymmetric p-modes trapped between $r_{\rm in}$ and 
the inner Lindblad resonance radius $r_{\rm IL}$ 
can grow due to corotational wave absorption, when 
the effective vortensity, $\zeta_{\rm eff}=\zeta/S^{2/\Gamma}$,
has a positive slope at the corotation radius.
As in the case of barotropic discs (Tsang \& Lai 2008; Lai \& Tsang 2009),
the general relativistic effect, where $\kappa^2$ is non-monotonic and 
becomes smaller as $r$ decreases toward $r_{\rm ISCO}$, plays a crucial
role in the instability. Now for non-barotropic discs, the entropy gradident
also plays an important role (cf.~Lovelace et al.~1999).

Our calculations of the global p-modes for various disc models (see
Table 1) indicate that the $m=3$ and $m=2$ modes (of lowest radial
order) have frequency ratio in the range of 1.57--1.69, similar to the
approximate 3:2 ratio as observed in high-frequency QPOs of black-hole
X-ray binaries (Remillard \& McClintock 2006).
The growth rates for these modes are significantly higher
than for the corresponding barotropic case (see Lai \& Tsang 2009), due
to the effect of the entropy in the effective vortensity. 
Although higher-$m$ modes may grow slightly faster, they would be less likely
to be observed due to averaging out of the luminosity variation over
the observable emitting area. The $m=1$ mode is found to have
a significantly smaller growth rate than the $m>1$ modes.

For our simple $\alpha$-disc models, we find that the p-mode
frequencies decrease slightly (by about 10-20\%) as the mass accretion
rate increases by a factor of 3. Observationally, it is known that
high-frequency QPOs are observed only when the X-ray binary systems
reside in the so-called steep power-law spectral state (also called
``very high state''), which may corresponds to a very specific range
of accretion rates. It is unclear whether our result is consistent
with the observed trend in high-frequency QPOs (e.g., Remillard et
al. 2002; Remillard \& McClintock 2006). Clearly, more
sensitive observations (e.g., with future X-ray timing missions;
see Barret et al.~2008, 
Tomsick et al.~2009)
would be useful to determine this trend and to search for the possible 
$m=4$ mode and the frequency ratios.

Finally, it should be noted that our calculations of global disc modes
are still based on rather crude models. The $\alpha$-discs are
phenomenological models, and our results (especially the mode growth rates) 
depend sensitively on the inner disc boundary conditions (both the
$l_0$ parameter for the background disc and the reflecting boundary
condition for the waves). 
Other potentially important effects (such as turbulence) have not been taken
into account. Thus we should treat our specific
results (such as those presented in Table 1) only as a demonstration of
the basic physical principles, and any comparison with the observations
at this point should be taken in this spirit.



\section*{Acknowledgments}
We thank Richard Lovelace and Michel Tagger for useful discussions
while we worked on this and related subjects during the last year or so.
This work has been supported in part by NASA Grant NNX07AG81G and NSF
grants AST 0707628.

\end{document}